\documentstyle[12pt]{article}

\begin{document}
\title{$\Upsilon(1s)\rightarrow\gamma(\eta',\eta)$ Decays}
\author{Bing An Li\\
Department of Physics and Astronomy, University of Kentucky\\
Lexington, KY 40506, USA}

\maketitle
\begin{abstract}
The decays of $\Upsilon(1s)\rightarrow\gamma(\eta',\eta)$ are studied by an approach which has successfully
predicted the ratio $\frac{\Gamma(J/\psi\rightarrow\gamma\eta')}{\Gamma(J/\psi\rightarrow\gamma\eta)}$.
Strong dependence on quark mass has been found in 
the decays $(J/\psi, \Upsilon(1s))\rightarrow\gamma(\eta',\eta)$. 
Very small decay rates of $\Upsilon(1s)\rightarrow\gamma(\eta',\eta)$
are predicted.

\end{abstract}
\newpage

Recently, CLEO Collaboration has reported measurements[1] of the branching ratios of radiative decays of 
$\Upsilon(1s)$ into $\eta$ and $\eta'$
\begin{eqnarray}
B(\Upsilon(1s)\rightarrow\gamma\eta)&<&1.0\times10^{-6}, \nonumber \\
B(\Upsilon(1s)\rightarrow\gamma\eta')&<&1.9\times10^{-6}
\end{eqnarray}
at the $90\%$C.L.  In an early search for $\Upsilon(1s)\rightarrow\gamma\eta'$ by CLEO[2] no signal 
in this mode has been found and in a $90\%$ confident level the upper limit of the branching ratio 
is about $1.6\times10^{-5}$. The previous 
CLEO search of $\Upsilon(1s)\rightarrow\gamma\eta$ produced an upper limit of $2.1\times10^{-5}$ 
at the $90\%$ confidence level[3]. Comparing with $B(J/\psi\rightarrow\gamma\eta'(\eta))$,
the branching ratios of $\Upsilon(1s)\rightarrow\gamma(\eta',\eta)$ are very small.

There are rich gluon physics in the radiative decays of heavy vector mesons to $\eta'$.
In QCD these processes are described by $V\rightarrow\gamma gg,\;\;gg\rightarrow\eta'$.
It is known for a long time that the coupling between $\eta'$ meson and two gluons is strong. 
The U(1) anomaly[4] of $\eta'$ meson is the first evidence.  
The quark components of $\eta'$ meson contribute only about ${1\over3}$ of $m_{\eta'}$, therefore, 
about ${2\over3}$ of $m_{\eta'}$ comes from gluon components of $\eta'$[5].
Experimental data[6] show that $B(J/\psi\rightarrow\gamma\eta') > B(J/\psi\rightarrow\gamma\eta)$
and $BR(J/\psi\rightarrow\omega(\phi)\eta) > BR(J/\psi\rightarrow\omega(\phi)\eta')$. 
These results support that $\eta'$ contains significant components of gluons[5].   
Now very small upper limit of $B(\Upsilon\rightarrow\gamma\eta')$ has been reported by CLEO[1].
The branching ratio is smaller
than $BR(J/\psi\rightarrow\gamma\eta')=(4.71\pm0.27)\times10^{-3}$ by almost three order of magnitudes.
This new channel can be used to study the application of QCD to the radiative decays of $\Upsilon$ and
the gluon content of $\eta'$.

The difference between the radiative decays of $J/\psi$ and $\Upsilon(1s)$ is caused by the mass difference of c- and b-quark.
As pointed out in Ref.[1] that the $B(\Upsilon(1s)\rightarrow\gamma\eta')$ predicted by the naive scaling is too large.
There are different theoretical approaches in the study of the branching ratios of 
$\Upsilon(1s)\rightarrow\gamma(\eta', \eta)$. 
Very small branching ratios for $\Upsilon\rightarrow \gamma\eta, \gamma\eta'$
($10^{-6}-10^{-7}$) are obtained by employing the Vector Meson Dominance Model in Ref.[7].     
The branching ratios of quarkonium decays to $\gamma\eta(\eta')$ $\sim(5-10)\times10^{-5}$, obtained 
in a nonrelativistic quark model[8] are larger than the upper limits(1). 
By using QCD sum rule and the U(1) anomaly, 
$B(\Upsilon\rightarrow\gamma\eta')=3.3\times10^{-5}$
and $B(\Upsilon\rightarrow\gamma\eta)=4.4\times10^{-6}$ are obtained in Ref.[9]. 
They are larger than the upper limits(1).
It is pointed out by the author of Ref.[10] that using QCD sum rule 
and the U(1) anomaly[9],  
the decay width of $V\rightarrow\gamma\eta'$ has a factor of ${1\over m^4_q}$($q=c, b)$ 
for $J/\psi$ and $\Upsilon$ respectively. This factor is the reason of larger $B(\Upsilon\rightarrow\gamma\eta')$ 
obtained in Ref.[9].
In Ref.[10] strong dependence on quark mass has been found. 
Using $m_\Upsilon\sim 2m_b$ and $m_{J/\psi}\sim 2m_c$, it is obtained  
\[B(\Upsilon(1s)\rightarrow\gamma+\eta')/B(J/\psi\rightarrow\gamma+\eta')\sim1.31(Q_b^2 m^6_c)/Q_c^2 m^6_b)
(\alpha(m_c)/\alpha_s(m_b)),\]
\[B(\Upsilon(1s)\rightarrow\gamma\eta)\sim 3.3\times10^{-7}, \; \;\;
B(\Upsilon(1s)\rightarrow\gamma\eta')\sim 1.7\times10^{-6}.\]
They are consistent with the upper limits(1).

In 1984 we have studied $J/\psi\rightarrow\gamma\eta, \gamma\eta'$[11].
In this short note we use the same approach to study $\Upsilon\rightarrow\gamma\eta'(\eta)$.
A brief review of the study done in Ref.[11] is presented below.
The decays are described as $J/\psi\rightarrow\gamma+g+g$ and 
two gluons are coupled to $\eta, \eta'$ respectively.  
The decay amplitude of 
$J/\psi\rightarrow\gamma+g+g$ is calculated by pQCD. $\eta$ and $\eta'$ have gluon components 
\begin{equation}
<G|T\{A^a_\alpha(x_1)A^b_{\beta}(x_2)\}|0>=\frac{\delta_{ab}}{\sqrt{2E_G}}
\epsilon_{\alpha\beta\mu\nu}(x_1-x_2)^\mu p^\nu f_G(x_1-x_2)e^{{i\over2}p_G (x_1+x_2)},
\end{equation}
where G is a gluon state with quantum number of $0^{-+}$. The function  $f_G(x_1-x_2)$ is
unknown. In Ref.[11] both a harmonic function with radius of one Fermi and  $f_G(0)$ have been tried.
The difference is about $10\%$. 
The possible dependence of $f_G(x_1-x_2)$ on $x_1-x_2$ has been ignored and $f_G$ is taken 
as a parameter. The decay widths are derived as
\begin{eqnarray}
\lefteqn{\Gamma(J/\psi\rightarrow\gamma\eta')=cos^2\theta sin^2\phi\frac{2^{11}}{81}\alpha 
\alpha^2_s(m_c)\psi^2_J(0)f^2_G{1\over m^8_c}\frac{(1-{m^2_{\eta'}\over m^2_J})^3}{\{1-2\frac{m^2_{\eta'}}{m^2_J}
+{4m^2_c\over m^2_J}\}^2}}
\nonumber \\
&&\{2m^2_J-3m^2_{\eta'}(1+{2m_c\over m_J})-16{m^3_c\over m_J}\}^2,\\
&&\Gamma(J/\psi\rightarrow\gamma\eta)=sin^2\theta sin^2\phi\frac{2^{11}}{81}\alpha
\alpha^2_s(m_c)\psi^2_J (0)f^2_G{1\over m^8_c}\frac{(1-{m^2_\eta\over m^2_J})^3}
{\{1-2\frac{m^2_\eta}{m^2_J}+{4m^2_c\over m^2_J}\}^2}
\nonumber \\
&&\{2m^2_J-3m^2_\eta(1+{2m_c\over m_J})-16{m^3_c\over m_J}\}^2,
\end{eqnarray}
where $\theta$ is the mixing angle between $\eta$ and $\eta'$ and $\phi$ is the mixing angle between 
a $0^{-+}$ flavor singlet and a $0^{-+}$ glueball, 
$\alpha_s(m_c)={g^2\over 4\pi}$ and g is the coupling constant of QCD, and
$\psi_J (0)$ is the wave function of $J/\psi$ at the origin, which is determined by the decay rate of
$J/\psi\rightarrow ee^+$
\begin{equation}
\psi^2_J(0)=\frac{27}{64\pi\alpha^2}m^2_J\Gamma_{J/\psi\rightarrow ee^+}.
\end{equation}
\(m_c=1.25\pm0.09\)GeV are presented in Ref.[6]. In the study of the amplitudes of $J/\psi\rightarrow
\gamma+f(1273)$[12] it is found that $m_c=1.3GeV$ fits the data very well and this value is  
in the range of $m_c$[6]. Taking this value of $m_c$ and \(\theta=-11^0\),
it is predicted[11]
\begin{equation}
\frac{\Gamma(J/\psi\rightarrow\gamma\eta')}{\Gamma(J/\psi\rightarrow\gamma\eta)}=5.1,
\end{equation}
which agrees with current experimental value $4.81(1\pm0.15)$[6].

Eqs.(3,4) show strong dependence of the decay rates on $m_c$.  
This effect originated in the pQCD calculation of $J/\psi\rightarrow\gamma+g+g$ and the $0^{-+}$ gluon components 
of $\eta,\; \eta'$(2). If \(m_J=2m_c\) is taken, the decay rates depend on ${1\over m^8_c}$. 
However, the value \(m_c=1.25\pm0.09\)GeV[6] shows that \(m_J=2m_c\) is not a good approximation.
Because of the cancellation between $m_J$ and $m_c$ in the factor of Eqs.(3,4)
\[\{2m^2_J-3m^2_{\eta'}(1+{2m_c\over m_J})-16{m^3_c\over m_J}\}^2,\;\;\;\{2m^2_J-3m^2_{\eta}(1+{2m_c\over m_J})-16{m^3_c\over m_J}\}^2\]
the decay rate is very sensitive to the value of $m_c$. This sensitivity has been found in the study of the decay 
$J/\psi\rightarrow\gamma+f(1273)$[12] too. In the right range of $m_c$[6] theory agrees with data very well.

By changing corresponding quantities and the electric charge of the quark, 
the decay rates of $\Upsilon(1s)\rightarrow\gamma\eta', \gamma\eta$ are 
obtained from Eqs.(3,4). The ratio is determined as
\begin{eqnarray}
\lefteqn{R_{\eta'}=\frac{B(\Upsilon\rightarrow\gamma\eta')}{B(J/\psi\rightarrow\gamma\eta')}}\nonumber \\
&=&
{1\over4}\frac{\alpha^2_s(m_b)}{\alpha^2_s(m_c)}\frac{\psi^2_\Upsilon (0)}{\psi^2_J (0)}{m^8_c\over m^8_b}
\frac{(1-\frac{m^2_{\eta'}}{m^2_\Upsilon})^3}{(1-\frac{m^2_{\eta'}}{m^2_J})^2}
\frac{(1-2\frac{m^2_{\eta'}}{m^2_J}+4\frac{m^2_c}{m^2_J})^2}
{(1-2\frac{m^2_{\eta'}}{ m^2_\Upsilon}+4\frac{m^2_b}{ m^2_\Upsilon})^2}
\frac{\{2m^2_\Upsilon-3m^2_{\eta'}(1+{2m_b\over m_\Upsilon})-16{m^3_b\over m_\Upsilon}\}^2}
{\{2m^2_J-3m^2_{\eta'}(1+{2m_c\over m_J})-16{m^3_c\over m_J}\}^3}
\frac{\Gamma_{J/\psi}}{\Gamma_\Upsilon},
\end{eqnarray}
where 
\begin{equation}
\frac{\psi^2_{\Upsilon}(0)}{\psi^2_J(0)}=4\frac{\Gamma_{\Upsilon\rightarrow ee^+}}
{\Gamma_{J/\psi\rightarrow ee^+}}\frac{m^2_{\Upsilon}}{m^2_J}.
\end{equation}

Taking \(m_J=2m_c\) and \(m_\Upsilon=2m_b\), it is obtained
\begin{eqnarray}
\lefteqn{R_{\eta'}=\frac{B(\Upsilon\rightarrow\gamma\eta')}{B(J/\psi\rightarrow\gamma\eta')}=
\frac{\Gamma(\Upsilon\rightarrow\gamma\eta')/\Gamma(\Upsilon\rightarrow light\; hadrons)}
{\Gamma(J/\psi\rightarrow\gamma\eta')/\Gamma(J/\psi\rightarrow light\; hadrons)}
\frac{B(\Upsilon\rightarrow light\; hadrons)}{B(J/\psi\rightarrow light\; hadrons)}}\nonumber \\
&=&\frac{\alpha_s(m_c)}{\alpha_s(m_b)}({m_c\over m_b})^7\frac{1-{m^2_{\eta'}\over 4m^2_b}}
{1-{m^2_{\eta'}\over 4m^2_c}}\frac{B(\Upsilon\rightarrow light\; hadrons)}{B(J/\psi\rightarrow light\; hadrons)}
\frac{\Gamma_{\Upsilon\rightarrow ee}}{\Gamma_{J/\psi\rightarrow ee}}
=0.29\frac{\alpha_s(m_c)}{\alpha_s(m_b)}({m_c\over m_b})^7.
\end{eqnarray}
Comparing with the ratio obtained in Ref.[10], stronger dependence on quark masses and small coefficient
are obtained by this approach. 

As mentioned above that \(m_J=2m_c\) is not a good approximation. In Ref.[6] \(m_b=(4.7\pm0.07)\) GeV is listed.
\(m_\Upsilon=2m_b\) works well. In this note \(m_b=4.7\) GeV and \(m_c=1.3\)GeV are taken.
Because of strong dependence on quark mass
both $\Gamma(J/\psi\rightarrow\gamma\eta', \gamma\eta)$ and $\Gamma(\Upsilon\rightarrow\gamma\eta', \gamma\eta)$
are sensitive to the values of $m_c$ and $m_b$ respectively. 
Inputing $B(J/\psi\rightarrow\gamma\eta')$, from Eq.(7) it is obtained
\begin{equation}
B(\Upsilon\rightarrow\gamma\eta')=R_{\eta'} B(J/\psi\rightarrow\gamma\eta')
=1.04\times^{-7},
\end{equation}
\begin{equation}
B(\Upsilon\rightarrow\gamma\eta)=0.022B(\Upsilon\rightarrow\gamma\eta')=0.23\times10^{-8}.
\end{equation}
Both branching ratios are less than the experimental upper limit[1].

The approach used in Ref.[11] is extended to study the decays of $\Upsilon(1s)\rightarrow\gamma\eta'(\eta)$.
Very strong dependence of the decay rate on quark mass is revealed.
The ratio $\frac{\Gamma(J/\psi\rightarrow\gamma\eta')}{\Gamma(J/\psi\rightarrow\gamma\eta)}$ and very small 
branching ratios of $\Upsilon(1s)\rightarrow\gamma\eta'(\eta)$ are predicted. The predictions agree with the data
very well.


\begin{thebibliography}{40}
\bibitem{} S.B.Athar et. al., CLEO Collaboration, Phys. Rev. {\bf D76}, 072003(2007).
\bibitem{} S.Richichi et al., CLEO Collaboration, Phys. Rev.Lett. {\bf 87}, 141801(2001).
\bibitem{} G.Masek et al., CLEO Collaboration, Phys. Rev. {\bf 65}, 072002(2002).
\bibitem{} E.Witten, Nucl.Phys. {\bf B156},269(1979); G.Veneziano, Nucl. Phys. {\bf b159},213(1979).
\bibitem{} B.A.Li, Phys. Rev. {\bf D74}, 034109(2006).
\bibitem{} W.M.Yao et al., Particle Data Group, J.of Physics G, {\bf 33},1(2006).
\bibitem{} G.W.Intemann, Phys. Rev. {\bf D27},2755(1983).
\bibitem{} J.Korner et al., Nucl. Phys. {\bf B229},115(1983); J.H.Kuhn, Phys. Lett. {\bf B127}, 257(1983).
\bibitem{} K.T.Chao, Nucl.Phys., {\bf B317},597(1989); {\bf B335}, 101(1990).
\bibitem{} J.P.Ma, Phys. Rev. {\bf D65}, 097506(2002).
\bibitem{} H.Yu, B.A.Li, Q.X.Shen, and M.M.Zhang, Phys. Energ. Fortis et Phys. Nucl.(Chinese),{\bf 8},285(1984).
\bibitem{} Bing An Li and Q.X. Shen, Phys. Lett. {\bf 126B}, 125 (1983).
\end{thebibliography}
\end{document}